\documentstyle[12pt]{article}
\newcommand {\beq}{\begin{equation}}
\newcommand {\eeq}{\end{equation}}
\newcommand {\beqa}{\begin{eqnarray}}
\newcommand {\eeqa}{\end{eqnarray}}
\newcommand {\beqan}{\begin{eqnarray*}}
\newcommand {\eeqan}{\end{eqnarray*}}
\newcommand {\n}{\nonumber \\}

\newcommand {\Romannumeral}[1]{\uppercase\expandafter{\romannumeral#1}}

\newcommand {\ee}{\mbox{e}}

\newcommand {\dd}{\mbox{d}}

\newcommand {\del}{\partial}

\begin{document}
\setlength{\oddsidemargin}{0cm}
\setlength{\baselineskip}{7mm}

\begin{titlepage}
 \renewcommand{\thefootnote}{\fnsymbol{footnote}}
    \begin{normalsize}
     \begin{flushright}
                 KEK-TH-394\\
                 KEK Preprint 94-15\\
                 ICRR-Report-318-94-13\\
                 UT-Komaba/94-11\\
                 April 1994
     \end{flushright}
    \end{normalsize}
    \begin{Large}
       \vspace{1cm}
       \begin{center}
         {\Large $R^2$ Gravity in $(2+\epsilon)$--Dimensional
Quantum Gravity} \\
       \end{center}
    \end{Large}

  \vspace{10mm}

\begin{center}
           Jun N{\sc ishimura}\footnote
           {E-mail address : nishi@danjuro.phys.s.u-tokyo.ac.jp,
{}~JSPS Research Fellow.},
           Shinya T{\sc amura}\footnote
           {E-mail address : shinya@danjuro.phys.s.u-tokyo.ac.jp} {\sc and}
           Asato T{\sc suchiya}\footnote
           {E-mail address : tutiya@danjuro.phys.s.u-tokyo.ac.jp}\\
      \vspace{1cm}
        $\ast$ {\it National Laboratory for High Energy Physics (KEK),}\\
               {\it Tsukuba, Ibaraki 305, Japan}\\
        $\dagger$ {\it Institute for Cosmic Ray Research, University
                  of Tokyo,} \\
                 {\it Tanashi, Tokyo 188, Japan}\\
        $\ddagger$ {\it Institute of Physics, University of Tokyo,} \\
              {\it Komaba, Meguro-ku, Tokyo 153, Japan}\\
\vspace{15mm}

\end{center}
\hspace{5cm}

\begin{abstract}
\noindent We study $R^2$ gravity in $(2+\epsilon)$--dimensional quantum
gravity. Taking care of the oversubtraction problem in the conformal mode
dynamics, we perform a full order calculation of string susceptibility
in the $\epsilon \rightarrow 0$ limit.
The result is consistent with that obtained through Liouville approach.
\end{abstract}

\end{titlepage}
\vfil\eject



Two-dimensional quantum gravity \cite{Polyakov}
has been studied intensively through
Liouville theory \cite{DDK} and the matrix model \cite{MM}
these several years,
and indeed its progress has provided us with much insight into
quantum gravity and string theory.
There are two ways to proceed further; one is to go beyond the so-called
$c=1$ barrier and
the other is to explore higher dimensional quantum gravity.
As for the former, $R^2$ gravity, which has been studied for years
\cite{Yoneya,ENO}, can be considered as a possible trial.
Recently Kawai and Nakayama \cite{KN} have succeeded in treating it
in the framework of Liouville theory.
It has been discovered that the surface becomes locally smooth
allowing a central charge greater than one.
Although $R^2$ gravity is not acceptable as a realistic string theory
due to loss of positivity,
it is still worth studying
as a new type of universality class of two-dimensional quantum gravity.
In contrast to ordinary gravity, $R^2$ gravity is difficult to study in
the matrix model and so far we do not have any other results which can be
compared with theirs.
It seems, therefore, desirable to investigate the theory in other approaches.

As for exploring higher-dimensional quantum gravity,
numerical simulations provide a powerful tool;
yet we should develop analytic approaches at the same time.
One of the possibilities is $(2+\epsilon)$--dimensional quantum gravity
\cite{2+epsilon,KKN,KST}.
The formalism, however, contains some subtlety concerning the oversubtraction
problem in the conformal mode dynamics,
as was discovered by Kawai, Kitazawa and Ninomiya \cite{KKN}.
Considering the situation,
we feel that it is worth while acquiring a deeper insight into the formalism
by applying it to other theories.
In this letter, therefore, we study $R^2$ gravity
in $(2+\epsilon)$--dimensional quantum gravity.
We obtain results consistent with ref. \cite{KN} by taking the
$\epsilon\rightarrow 0$ limit
in the strong coupling regime.
This provides another success in treating $R^2$ gravity.

\vspace{1cm}

We define a $(2+\epsilon)$--dimensional system corresponding to $R^2$ gravity
by the following action,
\beqa
S=\frac{{\mu}^{\epsilon}}{G}\int \dd^Dx \sqrt{g}R&+&\frac{{\mu}^{\epsilon}}
{4m^2}\int \dd^Dx \sqrt{g} R^2+\Lambda{\mu}^{\epsilon}\int \dd^D x \sqrt{g}\n
&+&{\mu}^{\epsilon} \sum_{i=1}^{c} \int \dd^D x \sqrt{g} \frac{1}{2}g^{\mu \nu}
\partial_{\mu}f_i\partial_{\nu}f_i,
\eeqa
where $G$ is the gravitational constant, $\Lambda$ is the cosmological
constant, $f_i$ is the matter field, and $D=2+\epsilon$. $m$ is a parameter
with mass dimension, which corresponds to the inverse of the range controlled
by the $R^2$ term.
Since the above action contains higher derivatives, which is difficult to
deal with, we introduce an auxiliary field $\chi$ and replace the $R^2$ term
with
\beq
{\mu}^{\epsilon} \int \dd^D x \sqrt{g} (-iR\chi+m^2{\chi}^2).
\eeq

In the Appendix, we calculate the one-loop counterterms
for the generalized action
\beq
S={\mu}^{\epsilon} \int \dd^D x \sqrt{g} \left(\frac{1}{2}K(\chi)g^{\mu\nu}
\partial_{\mu}\chi\partial_{\nu}\chi+L(\chi)R+M(\chi)\right)
+(\mbox{matters}),
\eeq
which reduces to the action considered by
setting $K(\chi)=0$, $L(\chi)=\frac{1}{G}-i\chi$ and
$M(\chi)=\Lambda+m^2{\chi}^2$.
As can be seen in (\ref{ctf}) and (\ref{P}), the counterterm for the
$\chi$--kinetic term can be set to $0$ by choosing appropriately the
function $f$, which comes from the freedom of
gauge fixing (\ref{eq:gaugefixing}).
Note also that the renormalization of $M(\chi)$ is self-contained,
which enables us to treat it separately as an inserted operator.
Thus the action including the one-loop counterterm reads
\beq
S+S_{\mbox{\scriptsize c.t.}}
={\mu}^{\epsilon} \int \dd^D x \sqrt{g} \left(\frac{1}{G}
-i\chi-\frac{1}{2\pi\epsilon}\frac{24-c}{12} \right) R +(\mbox{matters}).
\eeq
Special care should be taken for the counterterm
\beq
-\frac{24-c}{24\pi\epsilon}\mu^{\epsilon}\int \dd^D x \sqrt{g}R.
\label{os}
\eeq
Let us expand the metric as
\beqan
g_{\mu\nu}=\delta_{\mu\rho}(\ee^h)^{\rho}_{\ \nu}\ee^{-\phi},
\eeqan
where $\phi$ is the conformal mode and $h$ is the traceless symmetric
tensor field. The counterterm (\ref{os}) then yields the one for the
kinetic term of the conformal mode
$ \frac{24-c}{24\pi} \frac{1}{4} {\partial}_{\mu}\phi{\partial}_{\mu}\phi$,
which is $O(1)$, while the divergent one-loop diagram for the $\phi$
two-point function gives $O(\epsilon)$ quantity. The situation is just
the same as in ref. \cite{KKN}.  The counterterm (\ref{os}) is therefore
an oversubtraction for the conformal mode, which forces us to incorporate this
counterterm in the tree-level action and redo the perturbative expansion with
the effective action
\beq
S_{\mbox{\scriptsize eff}}
={\mu}^{\epsilon} \int \dd^D x \sqrt{g} \left(\frac{1}{G}
-i\chi-\frac{1}{2\pi\epsilon}\frac{24-c}{12}\right) R +(\mbox{matters}).
\eeq
This amounts to redefining
the $L(\chi)$ as $L(\chi)=\frac{1}{G}-i\chi-
\frac{24-c}{24\pi\epsilon}$.

We show in the following that one can obtain results consistent with
ref. \cite{KN} in the $\epsilon
\rightarrow 0$ limit in the strong coupling regime, {\it i.e.}
$G \gg \epsilon$. This is
to be expected, since in the infrared limit $R^2$ gravity reduces to ordinary
gravity without $R^2$ term,
which was reproduced in ref. \cite{KKN} also in the strong coupling regime.

\vspace{1cm}

Let us consider the renormalization of the operators $\int
\dd^{D}x\sqrt{g}$ and $\int \dd^{D}x \sqrt{g}\chi^{2}$.
We first show, up
to two-loop level, that the divergent parts coming from the diagrams with
$h_{\mu\nu}$ line cancel as a whole and therefore do not contribute
to the renormalization of the operators considered.

In order to diagonalize the kinetic terms in the action after gauge
fixing, we introduce the new quantum fields $\Phi, X$ and
$H^{\mu}_{~\nu}$ through
\beqa
   \phi & =& \tilde{F}\Phi + \frac{2L^{\prime}}
             {\epsilon L}\tilde{I}X  \nonumber\\
   \chi & = & \tilde{I}X \\
   h^{\mu}_{~\nu}& = & L^{-1/2}H^{\mu}_{~\nu},\nonumber
\eeqa
where $\tilde{I}$ and $\tilde{F}$ are given through
\beqan
\frac{1}{\tilde{I}^{2}} & = &
         \frac{{L^{\prime}}^{2}}{L}\left(1+\frac{D}{\epsilon}\right) \\
\tilde{F}^{2} & = & -\frac{4}{\epsilon DL}.
\eeqan
After this field redefinition, the kinetic term reduces to the
following standard form
\beq
  \int \dd^{D}x\sqrt{\hat{g}}\left\{
    \frac{1}{4}H^{\mu}_{~\nu,\rho}H^{\nu~~\!\rho}_{~\mu,}
    +\frac{1}{2}\hat{g}^{\mu\nu}\del_{\mu}\Phi\del_{\nu}\Phi
    +\frac{1}{2}\hat{g}^{\mu\nu}\del_{\mu}X\del_{\nu}X
    \right\}.
\eeq
The interaction vertices including $H_{\mu\nu}$ are
\beqa
&& \int \dd^{D}x\sqrt{\hat{g}}\left\{
  \frac{\epsilon}{4}(D-1)L^{1/2}\tilde{F}^2 H^{\mu\nu}
    \del_{\mu}\Phi\del_{\nu}\Phi
  -\frac{1}{\epsilon}(D-1)L^{-3/2}{L^{\prime}}^{2}\tilde{I}^{2}
    H^{\mu\nu}\del_{\mu}X\del_{\nu}X \right.  \nonumber \\
&& \hspace{2cm}\left. \mbox{}
    -\frac{\epsilon}{8}\tilde{F}^{2}H^{\mu\rho}H_{\rho}^{~\nu}
    \del_{\mu}\Phi\del_{\nu}{\Phi}
  +\frac{1}{2\epsilon}(D-1)L^{-2}{L^{\prime}}^{2}\tilde{I}^{2}
    H^{\mu\rho}H_{\rho}^{~\nu}\del_{\mu}X\del_{\nu}X+\cdots\right\}.
\eeqa
The operators can be written in terms of the new quantum fields as
\beqan
\int \dd^{D}x \sqrt{g} & = & \int \dd^{D}x \sqrt{\hat{g}}
    \mbox{e}^{-\frac{D}{2}(\tilde{F}\Phi +
                \frac{2L^{\prime}}{\epsilon L}\tilde{I}X)} \\
\int \dd^{D}x \sqrt{g}\chi^{2} & = & \int \dd^{D}x \sqrt{\hat{g}}
    \mbox{e}^{-\frac{D}{2}(\tilde{F}\Phi +
                \frac{2L^{\prime}}{\epsilon L}\tilde{I}X)}
              (\hat{\chi}+\tilde{I}X)^{2}.
\eeqan
The Figure shows the list of the diagrams with $H_{\mu\nu}$ line
we have to consider when we evaluate the one-point functions of the above
operators up to two-loop level. (a) and (b) correspond to $\langle
\Phi^{2} \rangle$, while (c) and (d) correspond to $\langle X^{2}
\rangle $.
Although each diagram has $O(\frac{1}{\epsilon})$ divergence ( Note
that $L \sim O(\frac{1}{\epsilon})$. ), an explicit calculation shows that
the divergent parts of (a) and (b), as well as (c) and (d), cancel
each other. One can also check that the contribution of the ghosts and
the matters is finite,
due to the suppression factors of $\epsilon$ and $L^{-1}\sim
O(\epsilon)$ in the action.
Thus we have shown that the diagrams containing $h_{\mu\nu}$, ghosts or
matters
do not affect the renormalization of the operators at least up to
two-loop level. We expect that this holds true to all orders of the loop
expansion and that two-dimensional $R^{2}$ gravity is completely
governed by the dynamics of the conformal mode $\phi$ and the
auxiliary field $\chi$.

Dropping the $h_{\mu\nu}$ field, the ghosts and the matters,
the effective action reads
\beqa
&&\int \dd^{D}x \sqrt{g}\left( \frac{1}{G} -i\chi -\frac{24-c}{24\pi\epsilon}
                                  \right) R \nonumber \\
& \sim & \int \dd^{D}x \sqrt{\hat{g}} \left\{
          -i\mbox{e}^{-\frac{\epsilon}{2}\phi}\left(\hat{R}
           -(D-1)\hat{g}^{\mu\nu}\nabla_{\mu}\del_{\nu}\phi
           +\frac{1}{4}\epsilon(D-1)\hat{g}^{\mu\nu}
              \del_{\mu}\phi\del_{\nu}\phi\right)
          \left( \chi +\hat{\chi}\right)\right. \nonumber \\
& & \hspace{3cm}\left. \mbox{}
+\left( \frac{1}{G} -\frac{24-c}{24\pi\epsilon} \right)
               \mbox{e}^{-\frac{\epsilon}{2}\phi}\left(\hat{R}
            -\frac{1}{4}\epsilon(D-1)\hat{g}^{\mu\nu}
              \del_{\mu}\phi\del_{\nu}\phi\right) \right\}.
\eeqa
Introducing new variables $\psi$ and $\xi$ through
\beqa
  \mbox{e}^{-\frac{\epsilon}{4}\phi}\chi & = & \xi \nonumber\\
  \mbox{e}^{-\frac{\epsilon}{4}\phi} & = &
                  1+\frac{\epsilon}{4}\psi,
\eeqa
the terms relevant to the renormalization of the operators considered
are
\beq
\sim \int \dd^{D}x \sqrt{g}\left\{
    i(D-1)\hat{g}^{\mu\nu}\del_{\mu}\psi\del_{\nu}\xi
   +\frac{24-c}{96\pi}(D-1)\hat{g}^{\mu\nu}
    \del_{\mu}\psi\del_{\nu}\psi \right\},
\eeq
which means that the problem is reduced to a free field theory with
the propagators
\beqa
\langle \psi(p)\psi(-p) \rangle & = & 0 \nonumber \\
\langle \psi(p)\xi(-p) \rangle & = & \frac{-i}{D-1}\frac{1}{p^2} \\
\langle \xi(p)\xi(-p) \rangle & = & \frac{24-c}{48\pi}
                   \frac{1}{D-1}\frac{1}{p^2}. \nonumber
\eeqa

Let us evaluate the divergence of the one-point functions of the
operators. As for the cosmological term, one gets
\beqan
\left\langle \int \dd^{D}x \sqrt{g}\right\rangle & = &
  \int \dd^{D}x \sqrt{\hat{g}}\left\langle \mbox{e}^{
      \frac{2D}{\epsilon}\log(1+\frac{\epsilon}{4}\psi)}
      \right\rangle \\
& = & \int \dd^{D}x \sqrt{\hat{g}},
\eeqan
due to $\langle\psi(p)\psi(-p)\rangle=0$. Thus one finds that the
cosmological term is not renormalized.
As for the mass term, one gets
\beqan
&&\left\langle \int \dd^{D}x \sqrt{g}\chi^{2}\right\rangle \\
& = &
  \int \dd^{D}x \sqrt{\hat{g}}\left\langle \left(
      \hat{\chi}^{2}+2\hat{\chi}\chi+\chi^{2}\right)\mbox{e}^{
         -\frac{D}{2}\phi}\right\rangle   \\
& = & \int \dd^{D}x \sqrt{\hat{g}}\left\{
\hat{\chi}^2+2i\frac{1}{2\pi\epsilon} \hat{\chi}
-\frac{18-c}{48\pi} \frac{1}{2\pi\epsilon}
-\left(\frac{1}{2\pi\epsilon} \right) ^2
\right\}  \\
& = & \int \dd^{D}x \sqrt{\hat{g}}\left\{\left(\hat{\chi}+
       \frac{i}{2\pi\epsilon}\right)^{2}
        -\frac{18-c}{48\pi}\frac{1}{2\pi\epsilon}\right\}.
\eeqan
Strictly speaking, one should have taken care of the $O(1)$
contributions to the term proportional to $\hat{\chi}$
in the last step of the equality.
One can check, however, that
starting from the action with $\chi$--linear term and adopting the minimal
subtraction scheme is equivalent to the above manipulation.

The bare operators, therefore, can be written as
\beqa
& & {m_{0}}^{2}\int \dd^{D}x \sqrt{g}{\chi_{0}}^{2}
      +\Lambda_{0} \int \dd^{D}x \sqrt{g} \n
&=& m^{2}\mu^{\epsilon}\int \dd^{D}x \sqrt{g}\left\{
      \left(\chi - i\frac{1}{2\pi\epsilon}\right)^{2}
      +\frac{18-c}{48\pi}\frac{1}{2\pi\epsilon}\right\}
      +\Lambda\mu^{\epsilon} \int \dd^{D}x \sqrt{g},
\eeqa
from which one can read off the relations between the bare parameters and
the renormalized ones as
\beqa
\chi_{0} & = & \chi-i\frac{1}{2\pi\epsilon} \nonumber \\
{m_{0}}^{2} & = & m^{2}\mu^{\epsilon} \\
\Lambda_{0} & = & \Lambda\mu^{\epsilon}+\frac{18-c}{48\pi}
    \frac{1}{2\pi\epsilon}m^{2}\mu^{\epsilon}. \nonumber
\eeqa
Using the above relations, one can evaluate the area dependence of the
partition function in the $\epsilon \rightarrow 0$ limit as follows.
\beqan
Z(A)=\int{\cal D}g_{\mu\nu}{\cal D}\chi_{0} \exp\left[-\mu^{\epsilon}
  \int \dd^{D}x\sqrt{g}\left( \frac{1}{G}-i\chi-\frac{24-c}{24\pi\epsilon}
\right)
  R-{m_{0}}^{2}\int \dd^{D}x\sqrt{g}\chi_{0}^{~2}
  -\Lambda_{0}\int \dd^{D}x\sqrt{g}\right]  &&\\
\cdot ~\delta\left( \left.\mu^{\epsilon}
     \int \dd^{D}x \sqrt{g} \right|_{\mu}-A \right).
    \hspace{5cm}&&
\eeqan
Rescaling the metric as $g_{\mu\nu} \rightarrow \lambda g_{\mu\nu}$,
\beqan
Z(A)&=&\int{\cal D}g_{\mu\nu}{\cal D}\chi_{0} \exp\left[
  -\lambda^{\epsilon/2}\mu^{\epsilon}
  \int \dd^{D}x
  \sqrt{g}\left(\frac{1}{G}-i\chi_{0}-\frac{12-c}{24\pi\epsilon}\right)
  R-\lambda^{D/2}m^{2} \mu^{\epsilon} \int \dd^{D}x
\sqrt{g}\chi_{0}^{~2}\right.\\
&& \mbox{}  -\left.\lambda^{D/2}\left( \Lambda\mu^{\epsilon}+\frac{18-c}{48\pi}
  \frac{1}{2\pi\epsilon}m^{2} \mu^{\epsilon} \right)
  \int \dd^{D}x\sqrt{g}\right]
 \cdot~ \delta\left( \lambda^{D/2}\left.\mu^{\epsilon}
  \int \dd^{D}x \sqrt{g} \right|_{\lambda^{1/2}\mu}
  -A\right) \\
&=&\int{\cal D}g_{\mu\nu} \exp\left[\frac{\epsilon}{2} \log{\lambda}
  \frac{12-c}{24\pi\epsilon} \mu^{\epsilon}\int \dd^{D}x
  \sqrt{g}R -\frac{\epsilon}{2}\log{\lambda}
  \frac{18-c}{48\pi}\frac{1}{2\pi\epsilon}m^{2}\lambda \mu^{\epsilon} \int
  \dd^{D}x \sqrt{g}\right] \\
&& \hspace{1cm} \cdot \exp\left[-\mu^{\epsilon}\int \dd^{D}x\sqrt{g}
\left(\frac{1}{G}R -\frac{12-c}{24\pi\epsilon}R +
\frac{1}{4m^{2}\lambda}R^{2}\right)
   -\Lambda_{0}\lambda\int \dd^{D}x \sqrt{g}\right] \\
&& \hspace{1.5cm}
\cdot ~\delta\left( \lambda^{D/2}\left.\mu^{\epsilon}
   \int \dd^{D}x \sqrt{g}\right|_{\mu}-A\right).
\eeqan
Setting $\lambda=A$,
\beqan
Z(A)&=& A^{\gamma_{\mbox{\tiny str}}-3}
    \mbox{e}^{-\Lambda_0 A}\int{\cal D}g_{\mu\nu}
    \exp\left[-\mu^{\epsilon}\int \dd^{D}x \sqrt{g}
    \left(\frac{1}{G}-\frac{12-c}{24\pi\epsilon}R
         +\frac{1}{4m^{2}A}R^{2}\right)\right]
        \\
&&  \hspace{3cm}\cdot ~\delta\left( \left.\mu^{\epsilon}
   \int \dd^{D}x \sqrt{g} \right|_{\mu}-1\right),
\eeqan
where $\gamma_{\mbox{\scriptsize str}}$ is the string susceptibility given by
\beq
\gamma_{\mbox{\scriptsize str}}
=2+\frac{c-12}{6}(1-h)+\frac{c-18}{192\pi^{2}}m^2 A.
\label{eq:strsus}
\eeq
For $m^2 A \ll 1$, the classical solution dominates in the path integral.
One gets, after taking account of the fluctuation around the classical
solution, the area dependence of the partition function as,
\beq
Z(A)\sim A^{\gamma_{\mbox{\tiny str}}-3}\mbox{e}^{-\Lambda A}\exp\left(-
   \frac{16\pi^{2}(1-h)^{2}}{m^{2}A}\right).
\eeq
One should note here that our $m^2$ correponds to $8\pi$ times the
$m^2$ of ref. \cite{KN}.
Comparing our result with that of ref. \cite{KN},
the only discrepancy is
the $c$--independent coefficient of $m^2 A$ in eq.(\ref{eq:strsus}),
which is subtraction scheme dependent.
We can, therefore, conclude that the two results are consistent.

\vspace{1cm}

In this letter we have studied $R^2$ gravity in the formalism of
$(2+\epsilon)$--dimensional quantum gravity.
 After taking care of the  oversubtraction problem and dropping the
$h$--field, the ghosts and the matters, the theory reduces to a free field
theory, which enables a full order calculation of the string susceptibility.
The result is consistent with that of ref \cite{KN}.
In our calculation, the peculiar $(c-12)$ factor comes from the shift
of the $\chi$--field and the $A$--dependent term comes from the fact
that the $\chi^2$ operator generates a cosmological term after
renormalization.

\vspace{1cm}

We would like to thank Prof. H. Kawai for stimulating discussion.
We are also grateful to Dr. A. Fujitsu, Dr. S. Ichinose,
Dr. N. Tsuda and Prof. T. Yukawa for providing us
with their results before publication.

\newpage

\section*{Appendix}
In this appendix, we calculate the one-loop counterterms for the most general
renormalizable action with a scalar field $\chi$ and $c$ species of conformal
matter,
\beq
S={\mu}^{\epsilon} \int \dd^D x \sqrt{g} \left(\frac{1}{2}K(\chi)g^{\mu\nu}
\partial_{\mu}\chi\partial_{\nu}\chi+L(\chi)R+M(\chi)\right)
+(\mbox{matters}).
\label{ga}
\eeq
Adopting the background field method, we replace $\chi$ with $\hat{\chi}+\chi$
and parametrize $g_{\mu \nu}$ as
\beqan
g_{\mu \nu}={\hat{g}}_{\mu\rho}(\ee^h)^{\rho}_{\ \nu} \ee^{-\phi},
\eeqan
where $\hat{\chi}$ and ${\hat{g}}_{\mu \nu}$ are the background fields, $\phi$
is the conformal mode, and $h$ is the traceless symmetric tensor field.
We expand the action up to the second order of $\chi,\ \phi$ and $h$, and drop
the first order terms following the prescription of the background field
method.
We can choose the gauge fixing term so that the mixing terms between
$h$ and the other fields may be cancelled,
\beq
S_{\mbox{\scriptsize g.f.}}={\mu}^{\epsilon} \int \dd^D x \sqrt{\hat{g}}
\frac{1}{2} L(\hat{\chi})\left(h^{\nu}_{\ \mu,\nu}+\frac{\epsilon}{2}
{\del}_{\mu} \phi -\frac{L'(\hat{\chi})}{L(\hat{\chi})} {\del}_{\mu} \chi-
f(\hat{\chi}) {\del}_{\mu} \hat{\chi} \chi\right)^2,
\label{eq:gaugefixing}
\eeq
where the comma represents the covariant derivative with respect
to ${\hat{g}}_{\mu \nu}$. Note that the function
$f$ can be taken arbitrary. The ghost action can be determined from
the gauge fixing term as
\beqa
S_{\mbox{\scriptsize ghost}}
&=&{\mu}^{\epsilon} \int \dd^D x \sqrt{\hat{g}} \left\{
  -\del_\nu {\bar{\eta}}^{\mu} \eta_{\mu,}^{\ \ \nu}-{\bar{\eta}}^{\mu}
  \hat{R}_{\mu}^{\ \nu} \eta_\nu+{\bar{\eta}}^{\mu}_{\ ,\mu}
  \frac{L'(\hat{\chi})}{L(\hat{\chi})}  {\del}_{\nu} \hat{\chi} \eta^\nu
  \right. \nonumber \\
&& \hspace{3cm}\left.+\left(\frac{L''(\hat{\chi})L(\hat{\chi})
   -{L'(\hat{\chi})}^2}
   {{L(\hat{\chi})}^2}
  -f(\hat{\chi})\right){\bar{\eta}}^{\mu} {\del}_{\mu} \hat{\chi}
  {\del}_{\nu} \hat{\chi} \eta^\nu \right\}.
\eeqa
The kinetic terms of $\chi, h$ and $\phi$, including those from the gauge
fixing term, thus read
\beqa
   \mu^{\epsilon} \int \dd^Dx \sqrt{\hat{g}}
     \left\{\frac{1}{2}\left(   K(\hat{\chi})
                            +  \frac{L'(\hat{\chi})^2}{L(\hat{\chi})}
                      \right) \hat{g}^{\mu\nu}\del_\mu\chi\del_\nu\chi
        +  \frac{1}{4}L(\hat{\chi})h^\mu_{\ \nu,\rho}h^\nu_{\ \mu,}\!{}^\rho
                                              \right. \nonumber\\
   \hspace{4cm}\left. +\frac{D}{2}L'(\hat{\chi})\hat{g}^{\mu\nu}
       \del_\mu\chi\del_\nu\phi
        -  \frac{\epsilon D}{8}\hat{g}^{\mu\nu}\del_{\mu}\phi\del_{\nu}\phi
    \right\}.
\eeqa
Note that the kinematical pole which appears in the case of ordinary gravity
does not show up here due to the $\phi$--$\chi$ coupling.
We introduce the new quantum fields $X, \Phi$ and $H^\mu_{\ \nu}$ through
\beqa
\chi&=&F(\hat{\chi})X-\frac{D}{2}F(\hat{\chi})^2L'(\hat{\chi})I(\hat{\chi})\Phi
                                                        \nonumber\\
\phi&=&I(\hat{\chi})\Phi\\
h^\mu_{\ \nu}&=&L(\hat{\chi})^{-1/2}H^\mu_{\ \nu}, \nonumber
\eeqa
where $F(\hat{\chi})$ and $I(\hat{\chi})$ are defined through
\beqa
\frac{1}{F(\hat{\chi})^2}
         &=&K(\hat{\chi})+\frac{L'(\hat{\chi})^2}{L(\hat{\chi})}\nonumber\\
\frac{1}{I(\hat{\chi})^2}
         &=&\frac{1}{4}\left(  \epsilon D L(\hat{\chi})
                             + D^2F(\hat{\chi})^2L'(\hat{\chi})^2
                       \right).
\eeqa
After this field redefinition, the kinetic term reduces to the standard form
\beq
\mu^{\epsilon} \int \dd^Dx \sqrt{\hat{g}}
             \left\{ \frac{1}{4}H^\mu_{\ \nu,\rho}{H^\nu_{\ \mu,}}^\rho
               -\frac{1}{2}\hat{g}^{\mu\nu}\del_\mu\Phi\del_\nu\Phi
               +\frac{1}{2}\hat{g}^{\mu\nu}\del_\mu X \del_\nu X
             \right\}.
\eeq
Taking account of the other terms coming from $S$,
$S_{\mbox{\scriptsize g.f.}}$ and $S_{\mbox{\scriptsize ghost}}$,
and calculating the one-loop counterterms through the
't Hooft-Veltman formalism \cite{tHV},
one obtains the final result for the action with the counterterms as
\beqa
S+S_{\mbox{\scriptsize c.t.}}&=&{\mu}^{\epsilon} \int \dd^D x \sqrt{g}
\left[\frac{1}{2}\left\{
K(\chi)-\frac{1}{2 \pi \epsilon}P(\chi)\right\}g^{\mu\nu}
\partial_{\mu}\chi\partial_{\nu}\chi \right.\nonumber  \\
&&\hspace{3cm}+\left\{L(\chi)-\frac{1}{2 \pi \epsilon}
  \frac{24-c}{12}\right\}R\n
&&\hspace{3cm}\left.+\left\{M(\chi)-\frac{1}{2 \pi \epsilon}
   \left(\frac{1}{2}M(\chi){I(\chi)}^2
+M'(\chi)\frac{1}{L'(\chi)}\right)\right\}\right]\nonumber \\
&&\hspace{3cm}+(\mbox{matters}),
\label{ctf}
\eeqa
where $P(\chi)$ is given by
\beqa
P&=&-2{L'}^{-2}L''+2L^{-1}L''+2F^2{L'}^{-1}L''(K'+2L^{-1}L'L''-L^{-2}{L'}^3)\n
 & &-\frac{1}{2}F^4{(K'+2L^{-1}L'L''-L^{-2}{L'}^3)}^2\n
 & &+3L^{-1}K-2L^{-1}F^{-2}+\frac{9}{2}L^{-2}{L'}^2-f.
\label{P}
\eeqa

\newpage
%

%
%
%
%
%
%
\newpage

\centerline{\Large Figure caption}
\bigskip
\noindent
The diagrams with $h_{\mu\nu}$ line we have to consider at
two-loop level.
The solid line, the dash line and the wavy line represent the propagators
of the $\Phi$--field, the $X$--field and the $H$--field respectively.
(a) and (b) correspond to $\langle \Phi^2 \rangle$, while
(c) and (d) correspond to $\langle X^2 \rangle$.
\\



\newpage

\begin{thebibliography}{99}

\bibitem{Polyakov} A.M. Polyakov, Mod. Phys. Lett. {\bf A2} (1987) 893.\\
                   V.G. Knizhnik, A.M. Polyakov
                  and A.B. Zamolodchikov, Mod. Phys. Lett. {\bf A3} (1988) 819.
\bibitem{DDK} F. David, Mod. Phys. Lett. {\bf A3} (1988) 1651.\\
              J. Distler and H. Kawai, Nucl. Phys. {\bf B321} (1989) 504.
\bibitem{MM} E. Brezin and V. Kazakov, Phys. Lett. {\bf B236} (1990) 144.\\
             M. Douglas and S. Shenker, Nucl. Phys. {\bf B335} (1990) 635.\\
             D.J. Gross and A.A. Migdal, Phys. Rev. Lett. {\bf 64} (1990) 717;
             Nucl. Phys. {\bf B340} (1990) 333.
\bibitem{Yoneya} T. Yoneya, Phys. Lett. {\bf B149} (1984) 111.
\bibitem{ENO} E. Elizalde, S. Naftulin and S.D. Odintsov, Int. J. Mod. Phys.
{\bf A9} (1994) 933.
\bibitem{KN} H. Kawai and R. Nakayama, Phys. Lett. {\bf B306} (1993) 224.
\bibitem{2+epsilon} S. Weinberg, in General Relativity, an Einstein Centenary
                    Survey, eds. S.W. Hawking and W. Israel (Cambridge
                    University Press, 1979).\\
                    R. Gastmans, R. Kallosh and C. Truffin, Nucl. Phys. {\bf
                    B133} (1978) 417.\\
                    S.M. Christensen and M.J. Duff, Phys. Lett. {\bf B79}
                    (1978) 213.\\
                    H. Kawai and M. Ninomiya, Nucl. Phys. {\bf B336} (1990)
115.
\bibitem{KKN} H. Kawai, Y. Kitazawa and M. Ninomiya, Nucl. Phys. {\bf B393}
(1993) 280; Nucl. Phys. {\bf B404} (1993) 684.
\bibitem{KST} S.-I. Kojima, N. Sakai and Y. Tanii, TIT preprint, TIT-HEP-238,
              November 1993, hep-th 9311045.\\
              Y. Tanii, S.-I. Kojima and N. Sakai, Phys. Lett. {\bf B322}
              (1994) 59.
\bibitem{tHV} G. 't Hooft and M. Veltman, Ann. Inst. Henri Poincare {\bf 20}
              (1974) 69.

\end{thebibliography}
\end{document}